\newcommand{\be}{\begin{equation}}
\newcommand{\ee}{\end{equation}}
\newcommand{\bea}{\begin{eqnarray}}
\newcommand{\eea}{\end{eqnarray}}
\newcommand{\ba}[1]{\begin{array}{#1}}
\newcommand{\ea}{\end{array}}
\documentclass[secnumarabic,amssymb,nobibnotes,nofootinbib,aps,pra,showpacs]{revtex4}
\usepackage{epsfig}
\usepackage{amsmath}
\usepackage{amssymb}
\usepackage{mathrsfs}
\usepackage{color}
\usepackage{rotating}
\usepackage{graphicx}
\usepackage{hyperref}
\usepackage{soul}

\begin{document}
\setlength{\topmargin}{-0.5in}

\title{Modeling atom-atom interactions at low energy by Jost-Kohn potentials}
\author{Subhanka Mal, Kingshuk Adhikary, Dibyendu Sardar, Abhik Kumar Saha and Bimalendu Deb}
\affiliation{School of Physical Sciences, Indian Association for the Cultivation of Science, Jadavpur, Kolkata 700032, India.}
\def\zbf#1{{\bf {#1}}}
\def\bfm#1{\mbox{\boldmath $#1$}}
\def\hf{\frac{1}{2}}
\begin{abstract}
 More than 65 years ago, Jost and Kohn [R. Jost and W. Kohn, {Phys. Rev.} {\bf 87}, 977 (1952)] derived an explicit expression for a class of short-range model potentials from a given effective range expansion with the $s$-wave scattering length $a_s$ being negative. For $a_s >0$, they calculated another class of short-range model potentials [R. Jost and W. Kohn, { Dan. Mat. Fys. Medd} {\bf 27}, 1 (1953)] using a method based on an adaptation from Gelfand-Levitan theory [I. M. Gel'fand and B. M. Levitan, { Dokl. Akad. Nauk. USSR} {\bf 77}, 557-560 (1951)] of inverse scattering. We here revisit the methods of Jost and Kohn in order  to explore the possibility of modeling resonant finite-range interactions at low energy. We show that the Jost-Kohn potentials can account for zero-energy resonances. The $s$-wave phase shift for positive scattering length is expressed in an analytical form as a function of the binding energy of a bound state. We show that, for small binding energy, both the scattering length and the effective range are strongly influenced by the binding energy; and below a critical binding energy the effective range becomes negative provided the scattering length is large. As a consistency check, we carry out some simple calculations to show that Jost-Kohn potentials can reproduce the standard results of contact interaction in the limit of the effective range going to zero. 
\end{abstract}
\maketitle
\section{Introduction}\label{1}
The purpose of this paper is to develop a description of $s$-wave resonant interactions between neutral particles at low energy in terms of the finite-ranged model interaction potentials derived in early fifties by Jost and Kohn \cite{JostKhon,JostKhon1}. Though there is a multitude of model potentials to describe physics of interacting particles at low energy at different length scales \cite{annphys:2008} there exists no unique or standard method to construct a model potential for the particles interacting through a scattering resonance with a finite range. Such a model potential for resonant interactions would be particularly important for many-body physics of ultracold atoms \cite{bloch:2008} with magnetically tunable Feshbach resonances  \cite{chin:rmp2010,sresonance} that make the atoms interact strongly. The well-known contact-type pseudo-potential approximation \cite{Huang1,Huang2} does not hold good in the case of resonant interactions with a large effective range. The strength of  contact potential is proportional to the $s$-wave scattering length $a_s$.  At a resonance, $a_s$ diverges, but this does not necessarily mean that mean-field interaction should diverge. The low density approximation $n |a_s|^3 <\!<1$ used in the case of a contact potential may breakdown for resonant interactions for which the effective range of interaction  becomes important. 

In recent times, several theoretical \cite{efferangetheo} and experimental works \cite{ohara:prl:2012,pra:2013:hulet,range} have demonstrated that the effective range at or near a Feshbach resonance becomes finite or large and even negative. Effective-range is shown to be quite important for three-body Efimov states \cite{efimov1,efimov2,efimov3,range,annphys:2012}. The finite-range and finite-energy effects of $s$-wave interaction has been shown to be incorporated in a contact-type pseudo-potential apprach with an energy-dependent scattering length
or phase-shift \cite{Bolda_02}.

Jost and Kohn constructed two classes of model potentials that include, among other parameters,  the effective range $r_0$ of interaction. One class is for negative $a_s$ \cite{JostKhon} and the other class is for positive $a_s$ \cite{JostKhon1}. The  model potential $V_{-}(r)$ ($r$ being the inter-particle separation) for  negative $a_s$ was derived by a perturbative inverse scattering method using the effective range expansion of the $s$-wave scattering phase shift. For positive $a_s$, the actual two-body interaction potential may support one or many bound states. Jost and Kohn had derived a class of ``equivalent" potentials $V_{+}(r)$ for positive $a_s$  from an analytical form of $s$-wave phase shift that includes the parameter $\kappa$ related to the binding energy $E_b = - \hbar^2 \kappa^2 /2\mu$ ($\mu$ is the reduced mass of the two particles and $\hbar$ is the Planck's constant divided by 2$\pi$). It is important to note that $V_{+}(r)$ does not support any bound state, but 
another ``equivalent" potential $V_b(r)$ can support the bound state with the same binding energy. $V_{+}(r)$ is a three parameter potential, the other two parameters are the $a_s$ and $r_0$ which are the same as corresponds to $V_b(r)$. It is a consequence of the theorem of Gel'fand and Levitan \cite{gelfand} that it is possible to construct a class of ``equivalent" potentials with the same phase shift but with or without bound state, showing independence of phase shifts from the bound states. However, from low energy scattering theory it follows that the positive $a_s$ may be related to the binding energy of a near-zero energy bound state. According to Gel'fand-Levitan theory, in order to construct $V_b(r)$ by an appropriate modification of $V_{+}(r)$, one has to include the normalization constant of the bound state that may be extracted from the asymptotic analysis of the scattering state through analytic continuation into the complex energy. Nevertheless, since both $V_{+}(r)$ and $V_{b}(r)$ are ``equivalent" yielding same scattering properties, one can work with $V_{+}(r)$ as far as elastic scattering properties of the system are concerned. 

Here we show that the Jost-Kohn potentials $V_{-}(r)$ and $V_{+}(r)$ are applicable to describe resonant interactions under certain physical conditions. They can naturally take into account the effective range effects of the interactions. We demonstrate that, in the limits $a_s \rightarrow \pm \infty$ and $\kappa \rightarrow 0$, both the potentials yield zero-energy resonance \cite{schieff_book, LandauLifshitz_QM}. We analyze in some detail how the tuning of the parameter $\kappa$ can control the value of the effective range. $V_{+}(r)$ is derived from an analytical form of the $s$-wave phase shift which is a function of $\kappa$. Note that the parameter $r_{0}$ and $a_{s}$ used to construct $V_{+}(r)$ correspond to the effective range and the scattering length only in the limit $\kappa\rightarrow\infty$. On the other hand, in the limit $\kappa\rightarrow 0$, the scattering length and the effective range become drastically modified due to the proximity of zero-energy resonance. As a result, the modified effective range may become large and negative. The Jost-Kohn potentials do not readily reduce to contact-type potentials in the limits $r_0 \rightarrow 0$ for small $|a_s|$. However, as a consistency check, we carry out numerical scattering calculations with Jost-Kohn potentials and show that in the limits $r_0 \rightarrow 0$, $\kappa \rightarrow \infty$ and for small $a_s$, the calculated results qualitatively reproduce the standard results that can be obtained from a contact interaction. Finally, we discuss in some detail how to fit Jost-Kohn potentials to describe Feshbach resonances under certain physical situations. In this context, it is worth mentioning that, recently several theoretical works \cite{theory-frmodel} have explored different procedures with a wide variety of model potentials to study the finite-range effects of low energy atom-atom interactions. For instance, Schneider {\it et al.} \cite{fr1} have used Born-Oppenheimer potential with an adjustable parameter to correspond to the experimental value of $a_s$, Lange {\it et al.} \cite{chin:pra:2009} have used a pair of square-well potentials with several adjustable parameters like binding energy and van der Waals length scale. Flambaum {\it et al.} \cite{Flambaum} have used a model Lennard-jones potential to explore finite-range effects near a Feshbach resonance. Veksler {\it et al.} \cite{Ketterle:pra:2014} have developed a modified inter-particle interaction to calculate corrections in the ground state solution to the Gross-Pitaevskii equation. The most widely used model is the two-channel model \cite{twochannel} with a pair potentials which depend on several experimental parameters of a particular system for which the Feshbach resonance is sought. Gao \cite{gao:jpb:2003} has given a prescription,based on his angular-momentum insensitive quantum defect approach \cite{gao:pra:2001}, how to construct model potentials of hard-sphere and Lenard-jones types with a $1/r^6$ asymptot. 

The paper is organized in the following way. In section \ref{2}, we analyze the method of construction of Jost-Kohn potentials. In section \ref{3}, we present and discuss our result showing the limits of zero-energy resonance and zero-range effects of Jost-Kohn potentials. We show that multi channel Feshbach resonances may be described by the Jost-Kohn potentials in some regimes. In the end, we conclude in section \ref{4}.  

\section{Jost-Kohn method}\label{2} 
Here we discuss the inverse scattering method of Jost and Kohn. Let us consider that a pair of particles interact via a spherically symmetric  potential $V(r)$ satisfying the condition $  \int_0^\infty |V(r)| r dr  <\infty$. The problem one addresses here is that, given the  phase shift $\eta_{\ell}(k)$ as a function of the wave number $k$ for a particular partial wave $\ell$,  whether it is possible to derive a model potential $V(r)$ that can reproduce the same $\eta_{\ell}(k)$. A treatise on this  problem was originally developed  by Gel'fand and Levitan \cite{gelfand}, and also by Jost and Kohn \cite{JostKhon,JostKhon1} who formulated a perturbative inverse scattering method.  It was first shown by Bargmann \cite{bargmann} and later corroborated by Jost and Kohn that one can  derive a class of equivalent potentials corresponding to the same phase shift. However, if there exists no bound state, then it is possible to derive a unique potential from the given function $\eta_{\ell}(k)$. Jost and Kohn obtained an explicit expression for a class of model potentials from the effective range expansion of $\eta_{0}(k)$ when there is no bound state and the $s$-wave scattering length $a_s$ is negative. Here we first  discuss the method of derivation of the negative-$a_s$ potential. Then we discuss the method to derive an equivalent potential for positive $a_s$.  
 \cite{JostKhon,JostKhon1}

The Schr\"{o}dinger equation of relative motion for $s$-wave is
\begin{equation}
 \frac{d^2\phi}{dr^2}+k^2\phi=U\phi
 \label{12}
\end{equation}
where  $U=2\mu V(r)/\hbar^2$, $k$ is  related to the collision energy $E=\frac{\hbar^2k^2}{2\mu}$ and $\mu$ is the reduced mass. 
Let $f(\pm k,r)$ be the two linearly independent solutions of Eq.(\ref{12}) with asymptotic boundary conditions
\begin{equation}
 \lim_{r \to \infty} e^{\mp ikr}f(\pm k,r)=1
 \label{13}
\end{equation}
A general solution $\phi(r)$ then asymptotically behaves as 
\begin{equation}
 \phi(r)\rightarrow f(-k)e^{-ikr}+f(k)e^{ikr}
\end{equation}
where $f(\pm k)$ = $f(\pm k,0)$ are called Jost functions. The scattering phase shift $\eta(k) = \eta_0(k)$ and $S$-matrix is given by 
\begin{equation}
 S(k)=e^{2i\eta(k)}=\frac{f(k)}{f(-k)}
 \label{14}
\end{equation}
and the Jost functions have the property $
 f(-k,0)=f^*(k,0)$.
Therefore, one obtains 
\begin{equation}
 \eta(k)={\rm Im} \left [ \log f(k) \right ]
 \label{1b}
\end{equation}
From Eq.(\ref{14}), one finds $
 \eta(k)+\eta(-k)=2n\pi$.  
If $n=0$ then $\eta(k)=-\eta(-k)$.

Using Green function, the solution $f(k,r)$ of Eq.(\ref{12}) can be expressed as an integral Volterra equation
\bea 
f(k,r) = e^{i k r} - \int_r^{\infty} k^{-1} \sin k(r'-r) U(r') f(k,r') d r' 
\label{volterra}
\eea 
In order to explore analyticity of scattering problem, let
$ z = 2 i k$, $ g(z,r) = g(2ik,r) = e^{-ikr}f(k,r)$. So,  
$g(z) = g(z,0)$ and $ \eta(k)={\rm Im} \left [\log g(z)\right ]$.
By multiplying both sides of Eq.(\ref{volterra}) by $e^{-i k r}$ and replacing $2i k$ by $z$,  one obtains 
\begin{equation}
 g(z,r)=1+\int_r^\infty\frac{1}{z}[1-e^{-z(r'-r)}]U(r')g(z,r')dr'
 \label{11}
\end{equation}
This equation can be solved by iteration  with the assumption ${\rm Re} [z]\geqq 0$. The function $g(z)$ is regular in ${\rm Re} [z]>0$ and continuous in ${\rm Re} [z]\geq 0$. After the successive iteration of Eq.(\ref{11}), we have 
\begin{widetext}
\begin{eqnarray}
g(z)-1&=&\sum_{l=1}^\infty \int_{0}^\infty dr_1 \int_{r_1}^\infty dr_2 ... \int_{r_{l-1}}^\infty dr_l\frac{1}{z^l} (1-e^{-zr_1})(1-e^{-z(r_2-r_1)})...(1-e^{-z(r_l-r_{l-1})})\nonumber\\ &\times& U(r_1)U(r_2)...U(r_l)
\label{55}
\end{eqnarray}
\end{widetext}
Under the approximation of small $U(r)$ this equation reduces to
\bea
g(z)-1\cong \frac{1}{z}\int_{0}^\infty dr (1-e^{-zr})U(r)
\eea
To the first approximation, $U(r)$ is replaced by an auxiliary potential $U_1(r)$ defined by 
\bea
g(z)-1=\frac{1}{z}\int_{0}^\infty dr (1-e^{-zr})U_1(r)
\eea
which can be recast into the form 
\bea 
\frac{1}{z} \left [ \frac{d}{d z} z g(z) - 1 \right ] = \int_{0}^\infty dr e^{-zr} U_1(r)
\eea 
So, $U_1(r)$ is given by the inverse Laplace transform of the function 
\bea 
\Phi_1(z) = \frac{1}{z} \left [ \frac{d}{d z} z g(z) - 1 \right ]
\eea 
Thus Eq.(\ref{55}) can be reformulated in the following form  
\begin{widetext}
\begin{eqnarray}
\int_{0}^\infty (1-e^{-zr})[U_1(r)-U(r)] dr &=& \sum_{l=2}^\infty \int_{0}^\infty dr_1 \int_{r_{l-1}}^\infty dr_l \frac{1}{z^{l-1}}(1-e^{-z(r_2-r_1)})...\nonumber\\&& \times (1-e^{-z(r_l-r_{l-1})})U(r_1)U(r_2)...U(r_l)
\label{laplace1}
\end{eqnarray}
\end{widetext}
The right hand side (RHS) of the above equation approaches zero as $z \rightarrow \infty$. This implies 
\begin{eqnarray}
\int_{0}^\infty [U_1(r)-U(r)] dr = 0
\end{eqnarray}
Thus $(U(r)-U_1(r))$ is given by the inverse Laplace transform of the RHS Eq.(\ref{laplace1}) in $z$. We can therefore write 
\bea 
U(r) = U_1(r) + \frac{1}{2 \pi i} \sum_{l=2}^{\infty} \int_{-i\infty}^{i \infty} e^{z r} \Phi_{l}(z) d z
\label{inverse}
\eea 
where 
\bea 
\Phi_l(z) &=& \int_{0}^\infty dr_1 \int_{r_{l-1}}^\infty dr_l \frac{1}{z^{l-1}}(1-e^{-z(r_2-r_1)})...\nonumber\\&& \times (1-e^{-z(r_l-r_{l-1})})U(r_1)U(r_2)...U(r_l)
\eea

The Eq.(\ref{inverse}) can be solved perturbatively. We next follow the Refs.\cite{JostKhon,JostKhon1} to elucidate how Jost and Kohn obtained  model finite-ranged potentials using effective range expansion.
\subsection{Negative-$a_s$ potential}
 
In the absence of any bound state, $\log g(z)$ becomes regular for ${\rm Re} [z]>0$ for $g(0)\neq 0$ and continuous for ${\rm Re}[z]\geq 0$,  its imaginary part being equal to the phase shift as given by $ \eta(k)={\rm Im} \left [\log g(z)\right ]$. Now, one can represent $\log g(z)$ in ${\rm Re}[z]\geq 0$ in terms of phase shift $\eta(k)$ by Poisson's integral 
\begin{equation}
 \log g(z)=-\frac{2i}{\pi}\int_{-\infty}^\infty \frac{\eta(k')}{2ik'-z}dk'
 \label{1h}
\end{equation}
Suppose, $\eta(k)$ admits an effective range expansion 
\bea
k\cot\eta(k)=-\frac{1}{a_s}+\frac{1}{2}r_{0}k^2+...
\label{etak}
\eea
where it is assumed that $a_s < 0$. Using the relation $\tan^{-1} x = (i/2) \log[(1-ix)/(1+ix)]$, one can express $\eta(k)$ in the form 
\bea 
\eta(k) = \frac{i}{2} \log \left [\frac{(z+a)(z-b)}{(z-a)(z+b)} \right ]
\label{etalog}
\eea 
where $a=\frac{2}{r_0}\left[ 1+\sqrt{1-\frac{2r_0}{a_s}}\right]$, $b=\frac{2}{r_0}\left[ -1+\sqrt{1-\frac{2r_0}{a_s}}\right]$. 
Substituting Eq.(\ref{etalog}) in Eq.(\ref{1h}), one obtains 
\begin{eqnarray}
g(z) &=& \frac{(z+b)}{(z+a)}=1+\frac{2a\lambda}{(z+a)}
\end{eqnarray}
where $\lambda= \frac{\left(b-a\right)}{2a}$  is a small parameter ($|\lambda|< 1$).  
Expanding the potential $U(r)$ in polynomial form
\begin{equation}
 U(r)= \sum_{m} \lambda^mU_m(r)
 \label{s1}
\end{equation}
each of the terms $U_m(r)$ can be calculated by inverse Laplace transform of $\Phi_m(z)$ as discussed above. The detailed derivation is given in the appendix-A. The resulting series 
can be expressed in a compact form, giving an explicit expression 
\bea 
U_{-}(r) &=& \frac{2a^2\lambda(1+\lambda)e^{-ar}}{[1+\lambda(1-e^{-ar})]^2}
\eea 
for negative $a_s$. From here onward, for the sake of simplicity, we consider $r_0$ as the unit of length, and the quantity $E_0=\hbar^2/(2\mu r_0^2)$ as the unit of energy, unless otherwise specified. 

\subsection{Positive-$a_s$ potential}
For positive scattering length, the potential may support bound states. The binding energies of the bound states given by the zeros $\xi_i$ of $g(z)$ in ${\rm Re}[z]>0$ lie on the real axis for $z$-plane. Therefore, in presence of bound states, the function $g(z)$ needs to be modified. Suppose, there exists $m$ bound states. Then the modified $g(z)$ reads as
\begin{equation}
 \bar g(z)=g(z)\prod_i^m\frac{(z+\xi_i)}{(z-\xi_i)}
 \label{eq25}
\end{equation}
which is non-zero for ${\rm Re}[z]>0$ and follow the same asymptotic properties of $g(z)$. The modified Eq.(\ref{1h}) have the form of 
\begin{equation}
 \log \bar g(z)=-\frac{2i}{\pi}\int_{-\infty}^\infty \frac{\bar \eta(k')}{2ik'-z}dk'
 \label{eq26}
\end{equation}
Let the binding energy of $i$ th  bound state energy be
\begin{equation}
 E_i =-\hbar^{2}\kappa_{i}^2/2\mu
\end{equation}

In deriving a potential that can support only one bound state and give positive $a_s$, Jost and Kohn first calculated an auxiliary potential $V_{+}(r)$ which may yield the same low energy scattering cross section but no bound state. The Jost function corresponding to $V_+(r)$ is assumed  to have the form  
\begin{equation}
\tilde{f}(k)=\frac{(2k-2i\kappa)^2}{(2k+ib)(2k-ia)}
 \label{f1}
\end{equation}
which obviously does not have a proper zero for the bound state. $V_+(x)$  is derived by iteration as above. This is expressed in terms of the three parameters 
which are $r_0$, $a_s$ and a dimensionless parameter $\Lambda$ related to the binding energy $E_b = -\hbar^{2} \kappa^2/2\mu$ of the bound state.    
The detailed method of derivation is discussed in appendix B. Explicitly, 
\begin{eqnarray}
V_{+}(x) &=& 8 e^{-2(1-\alpha) x}\frac{\Big[\big\{(1+\alpha\Lambda)(\alpha+\Lambda)(1-\alpha)(1-\Lambda^2e^{-2\beta x})\big\}^2- \Lambda^2\beta^2\big\{(1+\Lambda\alpha)^2e^{-2\alpha x} - (\alpha+\Lambda)^2e^{-2x} \big\}^2 \Big]}{\Big[(1+\alpha\Lambda)^2(\alpha+\Lambda^2e^{-2\beta x}) -(\alpha+\Lambda)^2(e^{-2(1-\alpha) x} + \alpha\Lambda^2 e^{-4x})\Big]^2}
 \label{1}
 \end{eqnarray}
where, $x=\frac{r}{r_0}$, $\alpha=\sqrt{1-\frac{2r_0}{a_s}}$, $\beta=1+\alpha$ and 
\bea 
\Lambda =\frac{\kappa r_0 -(1+\alpha)}{\kappa r_0 +(1+\alpha)}
\eea 
is a parameter with $-1<\Lambda<1$ and determined by $\kappa$.

\section{Results and discussion}\label{3}

We  calculate low energy elastic collisions for Jost-Kohn potentials using the well-known Numerov-Cooley algorithm \cite{nume1} to verify whether these potentials yield the same scattering length and the asymptotic states. The numerical results obtained for $V_{-}(r)$ agree absolutely well with any set of chosen parameters $a_s$ and $r_0$ used to construct $V_{-}(r)$.  For $V_{+}(r)$, the low energy scattering properties depend on the parameter $\kappa$ as we discuss below. We further show that, in the limits of $a_s \rightarrow \pm \infty$ and $\kappa \rightarrow 0$ or equivalently $\Lambda \rightarrow - 1$, the scattering solutions of Jost-Kohn potentials yield zero-energy resonances. On the other hand, in the limit  $r_0 \rightarrow 0$, $\kappa \rightarrow \infty $ or equivalently  $\Lambda \rightarrow 1$ and for $|a_s| < \infty$, the solutions of the potentials can reproduce the     
the standard results of zero-range or weak interaction. 

\subsection{Low energy expansion of the phase shift yielded by $V_{+}(r)$}
Though $V_{+}(r)$ does not support any bound state, it explicitly depends on the parameter $\kappa$ which determines binding energy $E_b$ of a bound state that is supported by 
an ''equivalent" potential, say $V_b(r)$. The $S$-matrix for $V_{+}(r)$ is $S = e^{2 i \tilde{\eta}} = \tilde{f}(k)/\tilde{f}(-k)$ where $\tilde{\eta}$ is the phase shift and $\tilde{f}(k)$ is given by Eq.(\ref{f1}). It is easy to deduce that in the limit $\kappa \rightarrow \infty$, $\tilde{\eta}$ becomes independent of $\kappa$ and so the phase shift in the low energy limit will be determined only by $a_s$ and $r_0$. On the other hand, 
 for small $\kappa$, $\tilde{\eta}$ will depend on all the three parameters $a_s$, $r_0$ and $\kappa$. As given by  Eq.(\ref{eq25}), we have
\begin{eqnarray}
\bar\eta(k)=\eta(k)-2\tan^{-1}\Big(\frac{\kappa}{k}\Big)
\label{eq29}
\end{eqnarray}
where $\eta(k)$ corresponds to $V_b(r)$ and is assumed to have an  effective range expansion in terms of $a_s$ and $r_0$ as in Eq. (18) but for $a_s >0$. The expression (\ref{eq29}) shows that, for $\kappa \ne 0$ and $k \rightarrow 0$, 
$\tilde{\eta} = \eta - \pi$. So, both the potentials $V_{+}(r)$ and $V_b(r)$ will yield the same $s$-wave scattering cross section at low energy. The question here is how $\kappa$ affects the effective range expansion. 

From Eq.(\ref{eq29}), we obtain  
\begin{eqnarray}
k\cot\bar\eta(k) &=& \frac{1}{\big(\frac{2}{\kappa}-a_s\big)}\frac{\bigg[1+\Big(2\kappa a_s-\frac{r_0a_s\kappa^2}{2}-1\Big)\frac{k^2}{\kappa^2}+\frac{r_0a_s}{2\kappa^2}k^4\bigg]}{\bigg[1+\Big(\frac{1-r_0\kappa}{\frac{2}{\kappa a_s}-1}\Big)\frac{k^2}{\kappa^2}\bigg]}   
 \label{de} 
\end{eqnarray}
Assuming that $\kappa a_s > 2$ or $\kappa a_s < 2$ and $k \ll \kappa $ (for low energy scattering), we have $1\gg  \mid  \frac{1-r_0\kappa}{\frac{2}{\kappa a_s}-1}   \mid \frac{k^2}{\kappa^2}$.
After binomial expansion in Eq.(\ref{de}), up to the second order in $k/\kappa$, we get
\begin{equation}
 k\cot\bar\eta(k) = -\frac{1}{\bar a_s}+\frac{1}{2}\bar r_{0}k^2+...
 \label{modifyeta}
\end{equation}
where
\begin{eqnarray}
\label{eq32}
\bar a_s &=& a_s-\frac{2}{\kappa} \\
\bar r_{0} &=& \frac{a_s}{\bar a_s} \Bigg ( \frac{1}{\kappa} \Bigg) \Bigg[ \kappa r_0-4 +\frac{1}{2\kappa a_s}+\frac{1- 2 r_0\kappa}{4- 4 \kappa a_s}\Bigg]
\label{eq33}
\end{eqnarray}
where $\bar r_{0}$ is the modified range. From this formula, it is clear that $\bar{r}_0 $ is negative but $\bar{a}_s >0$ for $\kappa r_0 <\!< 1$ and $\kappa a_s > 2$.  On the other hand,  $\bar{a}_s < 0$ for $\kappa a_{s} <2$. So, for $\kappa a_s <\!< 1$,   $\bar{r}_0$ may vary from  positive to negative values as $\kappa r_0$ changes 
from small ($<\!< 1$) to large values ($>\!> 1$).  The negativity of $\bar{r}_0$ may be interpreted as resulting from the breakdown of the standard effective range expansion due to the proximity of a  zero-energy resonance as we describe in the next subsection. In the standard effective range expansion \cite{bethe,LandauLifshitz_QM} as in Eq.(\ref{etak}), it is assumed that $kr_{0}<<1$ and $r_{0}>0$. For $V_{+}(r)$, we have found that the numerically calculated $a_s$ and $r_0$  agree quite well with the chosen values of $a_s$ and $r_0$ used to construct $V_{+}(r)$ if $\kappa r_0 > \!> 4$. We find that the numerical calculated values of $\bar{a_s}$ and $\bar{r_0}$  deviate substantially from the chosen $a_s$ and $r_0$ if $\kappa r_{0}<<1$.

\subsection{Resonant interactions}

In this subsection we first calculate the scattering phase shift and cross section as a function of collision energy $E$ or wave number $k$ for  Jost-Kohn potentials. Zero-energy resonance occurs when $\eta(0) = \pi/2$, that is, the $s$-wave phase shift at $k=0$ is $\pi/2$. This happens if $f(0)=0$, physically this implies that the potential is about to support a bound state at an energy given by  $f(k \ne 0) = 0$ if the potential is slightly modified.  This follows from the fact that there exists no bound state at zero energy for $s$ wave unlike that at higher partial waves. A bound state for $s$-wave can exist only at finite energy, in which case $f(0) \ne 0$ and  $\eta(0) = \pi$ \cite{schieff_book}.

\begin{figure}[h]
\centering
\vspace{.4in}
\begin{tabular}{@{}cccc@{}}
\hspace{-0.2in}
\includegraphics[height=2.45in, width=3in]{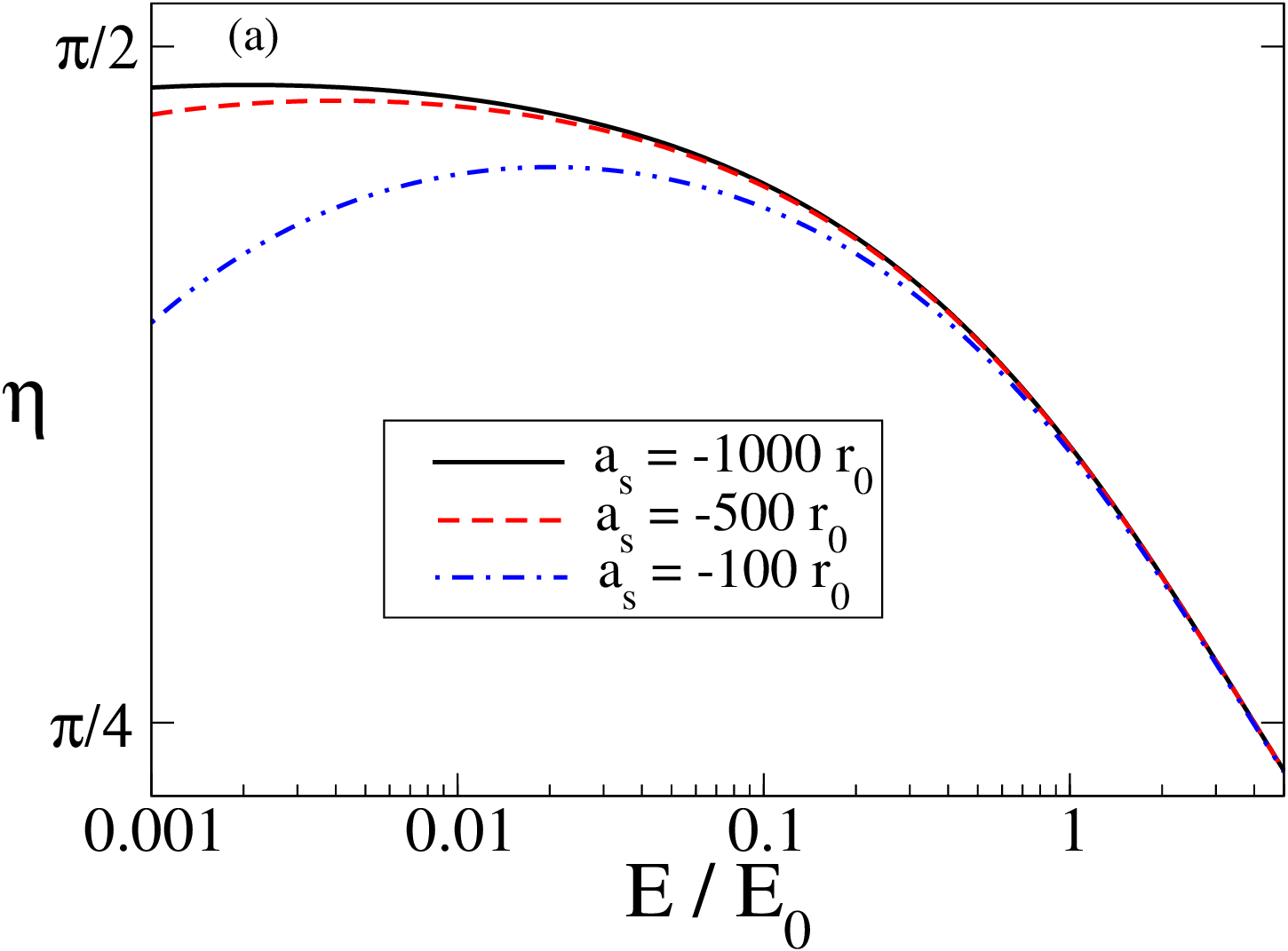} &
\includegraphics[height=2.5in, width=3in]{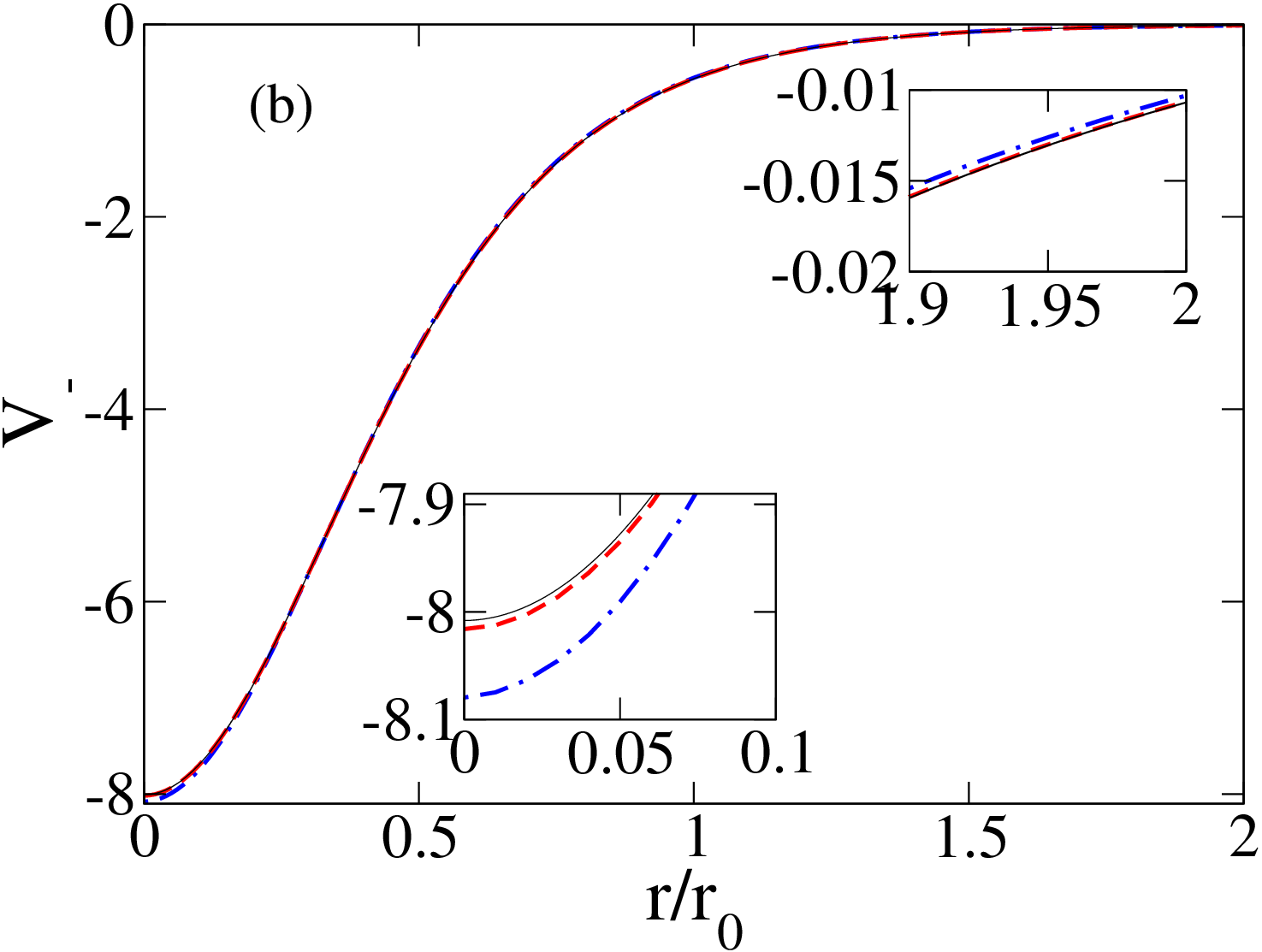}\\
\end{tabular}
\caption{(a) Variation of $s$-wave scattering phase shift $\eta$ as a function of dimensionless energy $E/E_{0}$ for different values of negative scattering length. (b) Variation of $V_{-}$ (in unit of $E_0$) as a function of iseperation in unit of $r_0$ for different values of negative $a_{s}$.} 
\label{negative-as}
\end{figure}

From the effective range expansion, it follows that for $a_s \rightarrow - \infty$, we have $\eta(0) \rightarrow \pi/2$. Since $V_{-}(r)$ is derived based solely on the effective range expansion, one would expect that numerically calculated phase shift for $V_{-}(r)$ should reproduce this result. In fact, the calculated phase shift as plotted in Fig.\ref{negative-as}  shows this expected behavior.

The Jost function corresponding to the potential $V_{+}(r)$ is given by $\tilde{f}(k)$ of Eq.(\ref{f1}). One can notice that $\tilde{f}(0) \ne 0$ if $\kappa \ne 0$.  In this case, in the limit $k \rightarrow 0$, the $S$-matrix $S(k) = \tilde{f}(k)/\tilde{f}(-k)$ approaches identity and so the the phase shift $\eta(0) \rightarrow \pi$. On the other hand, for $\kappa = 0$, $a_s= \infty$, we have $S(k \rightarrow 0) \rightarrow -1$. The expression Eq.(\ref{modifyeta}) shows that for $\kappa a_{s}<2$, $k\rightarrow 0$, $a_{s}\rightarrow\infty$ with $ka_{s}>>1$, we have  $\cot\bar\eta(k)\rightarrow 0_{+}$ implying that $\eta(k\rightarrow 0)=\pi/2$. Therefore the system exhibits zero-energy resonance.  
\begin{figure}[h]
\centering
\vspace{.4in}
\begin{tabular}{@{}cccc@{}}
 \hspace{-0.2in}
\includegraphics[height=2.45in, width=3in]{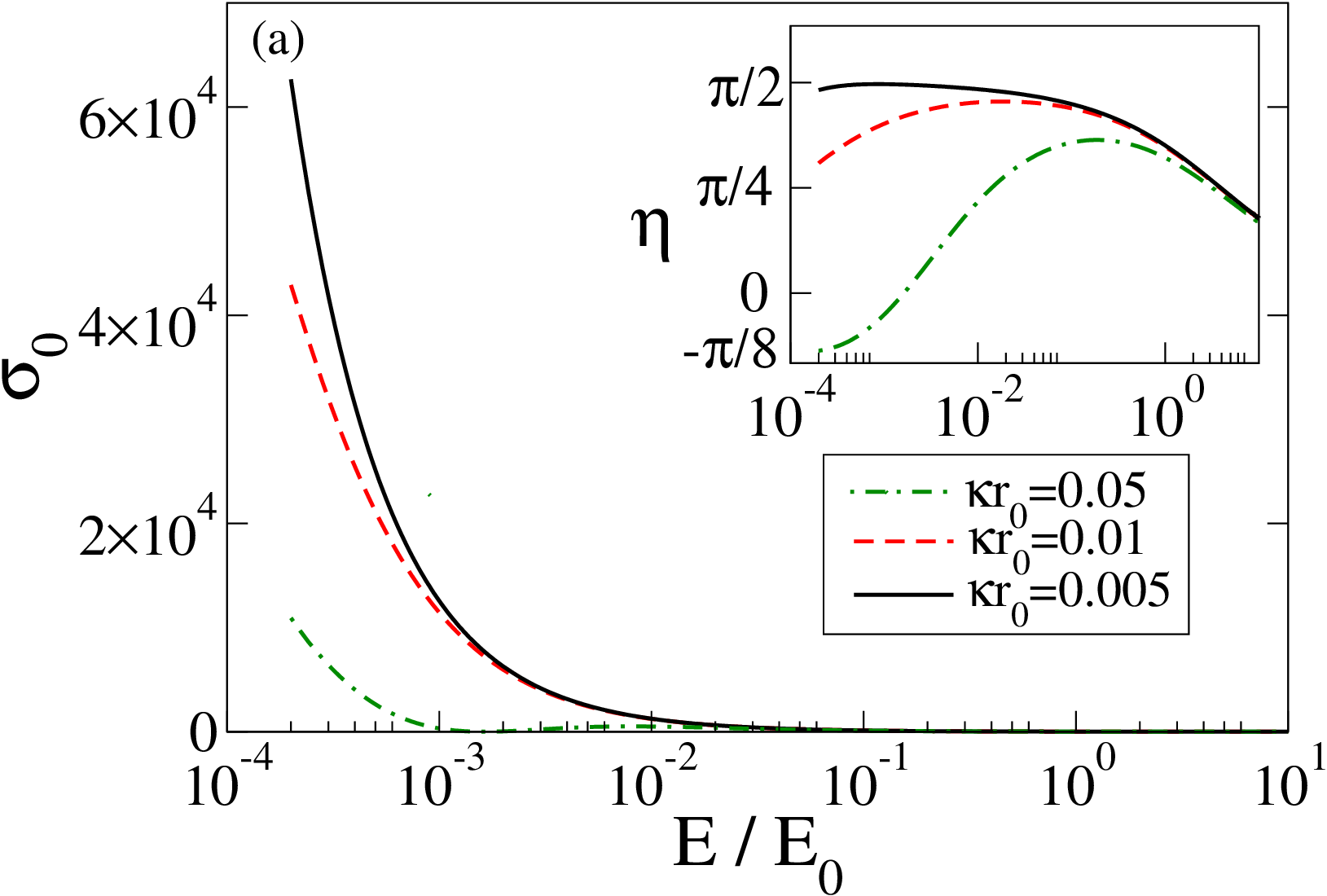} &
\includegraphics[height=2.5in, width=3in]{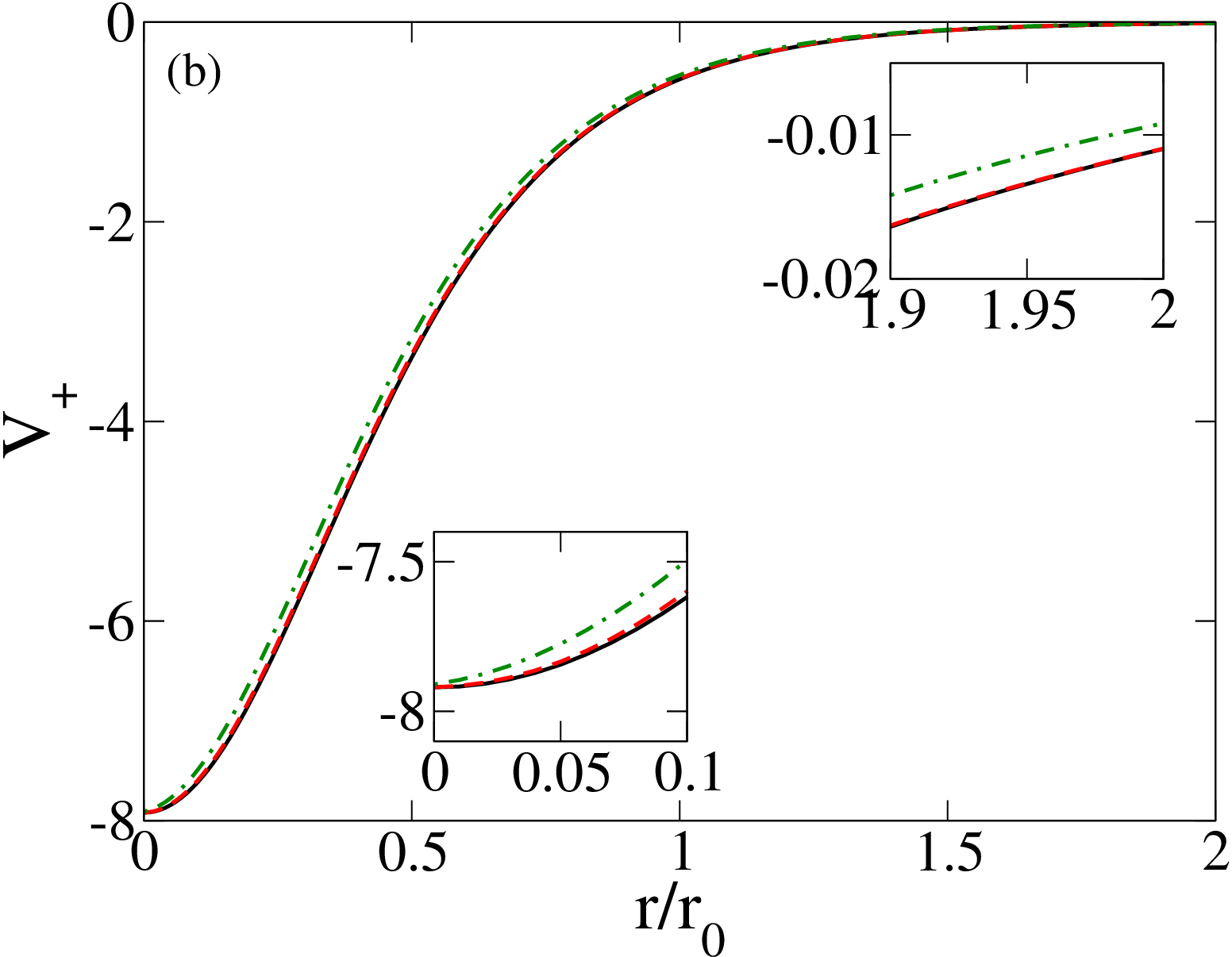}\\
\end{tabular}
\caption{\small (a) Variation of s-wave scattering cross-section $\sigma_{0}$ in unit of $r_{0}^2$ as a function of dimensionless energy $E/E_{0}$ for $a_{s}=100r_{0}$ for three different values of $\kappa=0.05r_{0}^{-1}$ (dashed-dotted), $\kappa=0.01r_{0}^{-1}$ (dashed) and $\kappa=0.005r_{0}^{-1}$ (solid). The corresponding phase shifts for different values of $\kappa$ are shown in the inset. (b) Variation $V_{+}$ (in unit of $E_0$) as a function of $r/r_0$ for different values of $\kappa$.}
 \label{Figure1}
\end{figure}
In Fig.\ref{Figure1}. we have plotted the $s$-wave scattering cross sections $\sigma_{0}(E)$ as a function of dimensionless energy $E/E_{0}$ in the resonance limit. Consequently $\sigma_{0}(E)$, shows a divergent signature in the limit $E\rightarrow 0$ as in Fig.\ref{Figure1}. From the inset of Fig.\ref{Figure1}, it is clear that the phase shift goes to $+\frac{\pi}{2}$ as $\kappa \rightarrow 0$. In this context, it is worth mentioning that both the potentials $V_{+}(r)$ and $V_{-}(r)$ reduce to the same analytical form which is of Potsch-Teller potential in the limits $\Lambda \rightarrow -1  $ (or $\kappa \rightarrow 0 $) and $a_s \rightarrow \pm \infty$ \cite{deb:injmp:2016}, signifying zero-energy resonance. It is worth mentioning that the same from of P\"{o}schl-Teller potential has been used earlier for quantum Monte Carlo simulation of many-body physics of an ultracold Fermi gas of atoms \cite{carlson:prl:2003,forbes:pra86:2012,li:pra84:2011,morris:pra81:2010,schonenberg:pra:95:2017}.
 
\subsection{Zero range limit of Jost-Kohn potentials}
\begin{figure}[h]
\centering
\vspace{.4in}
  \includegraphics[height=3in, width=3.5in]{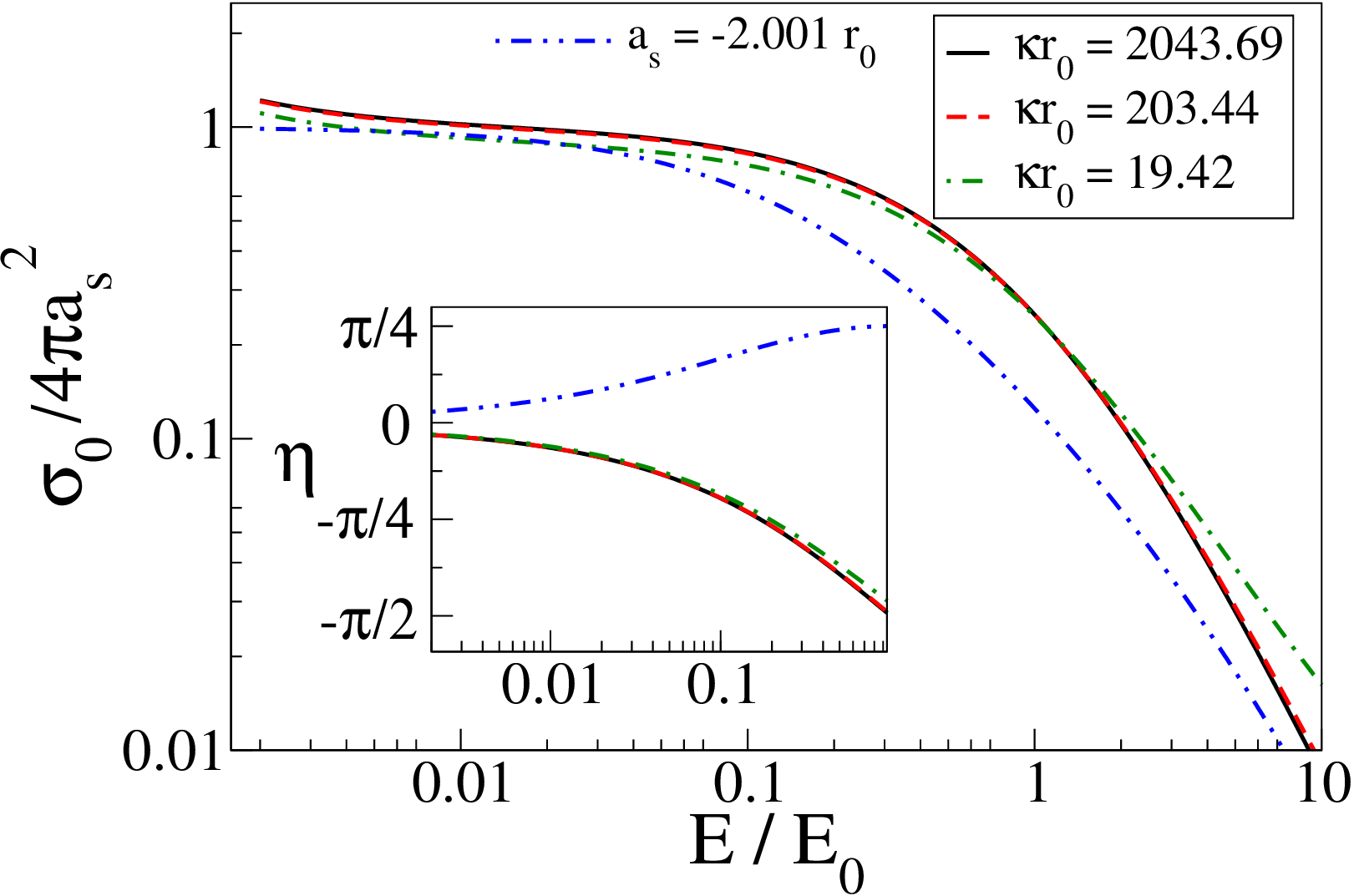}
 \caption{\small Variation of $\sigma_{0}$(in unit of $4\pi a_{s}^2$) as a function of $E/E_{0}$ in the almost zero-range limit considering both positive (black,red,green) and negative (blue) values of scattering length for a set of three higher values of $\kappa$. The corresponding phase shifts are shown in the inset.}
 \label{Figure2}
\end{figure}

Jost-Kohn potentials  do not explicitly reduce to a delta-function like zero-range contact  potential in the limit $r_0 \rightarrow 0$. The derivation of $V_{+}(r)$  makes use of the assumption $a_s>2r_0$. Here we numerically verify whether the Jost-Kohn potentials reproduce the known results of weak interaction regime (small $|a_{s}|$) in the limit $r_{0}\rightarrow0$. In Fig.\ref{Figure2}. we show the variation of scattering cross section  as a function of  energy, for large $\kappa$. Here, the quantity $\frac{\sigma_{0}}{4\pi a_{s}^2}$ nearly equals to unity at very low energy limit. We have verified this limit for different values of $\kappa$ and scattering lengths. The inset of Fig.\ref{Figure2} exhibits the behavior of the phase shift in the limit $E\rightarrow 0$. We notice that $\eta(E\rightarrow 0)\propto \pm \sqrt{E}$, where $+(-)$ corresponds to negative(positive) $a_{s}$. This low energy behavior of $\eta(E)$ is consistent with that for a contact or weak interaction potential.  

Having shown that the Jost-Kohn potentials can describe the standard low energy scattering properties of a pair of ultracold atoms in free space, we now discuss whether these potentials are good enough to model the interaction between a pair of trapped atoms. There exists an exact solution for a pair of ultracold atoms interacting via a regularized contact potential in a 3D isotropic harmonic oscillator \cite{butsch:foundphys:1998}. Several theoretical studies \cite{studies} have shown that this exact solution is good enough so long as $|a_s|$ is much smaller than the characteristic length scale or more specifically the size of the ground state of the isotropic harmonic oscillator. In a previous study \cite{physscrpt:partha}, it has been demonstrated that the bound state solutions of a pair of ultracold atoms interacting via  Jost-Kohn potentials in an isotropic harmonic oscillator can qualitatively reproduce the results of Ref.\cite{butsch:foundphys:1998} when $r_0$ is much smaller than the harmonic oscillator length scale provided $a_s $ is small enough. Also, for a quasi-one dimensional trap, Jost-Kohn potentials are shown to agree qualitatively with the results of a contact interaction only when $r_0$ is much smaller than the length scale of the transverse 2D harmonic oscillator and small $a_s$ \cite{physscrpt:partha}.

\subsection{Feshbach resonances}

Here we discuss to what extent and under what physical conditions it may be  possible to describe Feshbach-resonant interactions between ultracold atoms by Jost-Kohn potentials. Feshbach resonance is a multichannel scattering problem where a quasi-bound state supported by one or multiple closed channels is made degenerate or quasi-degenerate with the bare scattering state of at least one open channel by means of an external magnetic or optical field. As a consequence, a Feshbach resonance occurs due to an admixture of bound and continuum states leading to a dressed continuum which can also be dealt with Fano's method \cite{fano1961,debandgsa:jpb:2009}. If there is only one open channel, then the physical $S$-matrix derived upon elimination of all the closed channels corresponds to the open channel only. If the experimentally determined phase shift corresponding to this effective single-channel physical $S$-matrix element admits an effective range expansion with the effective range 
 $r_{f}$ (we use different notation for the effective range of the Feshbach resonance to distinguish it from $r_0$ that corresponds to Jost-Kohn potentials) then the method of Jost and Kohn will be definitely applicable to this effective single-channel problem. The pertinent question here is how to reduce a multi-channel scattering problem into an effective single-channel one describable by the Jost-Kohn potentials. 
 
 In the current literature on magnetic Feshbach resonances \cite{chin:rmp2010,sresonance} of ultracold atoms, the resonances are mainly categorized into two types depending on the width of the resonance: narrow or closed-channel dominated and broad or open-channel dominated ones.  The width of a Feshbach resonance is quantified by the dimensionless strength parameter defined by $s_{res}=\frac{\delta\mu \Delta}{\bar E}\frac{a_{bg}}{\bar a}$, where  $\delta\mu$ is the difference between the magnetic moments of the bare quasi-bound state and the two separated atoms, $a_{bg}$ is the background scattering length and $\Delta$ in unit of magnetic field strength is the width of the resonance. 
 For a potential behaving asymptotically as $-\frac{C_{6}}{r^{6}}$, with $C_{6}$ being the van der Waals' coefficient, Gribakin and Flambaum \cite{vanderwaal} defined a length scale called the mean scattering length $\bar a=\frac{2\pi}{\Gamma(1/4)^2}\Big(\frac{2\mu C_{6}}{\hbar^{2}}\Big)^{1/4}$, and corresponding energy scale $\bar E=\frac{\hbar^{2}}{2\mu\bar a^2}$.  A narrow or 
closed-channel-dominated resonance occurs when $s_{res}<1$
while for a broad or entrance-channel-dominated resonance,  $s_{res}>1$. Gao \cite{gao} and  Flambaum {\it et al.}\cite{Flambaum} have defined an effective range $R_e$ for asymptotic van der Waals potential 
by $R_e \approx \left(\frac{\Gamma(\frac{1}{4})^4}{6\pi^2}\right)\bar a\left[1-2\left(\frac{\bar a}{a_s}\right)+2\left(\frac{\bar a}{a_s}\right)^2\right]$. The $r_{f}$-dependence of a narrow Feshbach resonance has been shown to be quite different from that of a broad Feshbach resonance \cite{effrange}. It has been further demonstrated, both theoretically \cite{efferangetheo} and experimentally \cite{range,pra:2013:hulet} that $r_{f}$ near a narrow Feshbach resonance may become quite large, negative and magnetic field-dependent. The experimental observation \cite{range} shows that near the narrow Feshbach resonance ($B_0=58.9$G) of $^{39}$K the effective range sharply changes with scattering length (or magnetic field) and the effective range is found to be large negative in the vicinity of this Feshbach resonance. For the case of $^6$Li NFR near 543.3G, similar results are found \cite{pra:2013:hulet}. In contrast,  $r_{f}$ near a broad Feshbach resonance is usually positive, small and close to $R_e$. In the case of an intermediate-range Feshabch resonance ($s_{res} \sim 1$), quite interesting field-dependence of $r_{f}$ has been experimentally demonstrated \cite{pra:2013:hulet}. From the analysis made in subsection 3.2, we understand that the effective range of the resonant interaction with $a_s>0$  may be controlled by the parameter $\kappa$ of the potential $V_{+}$.

In the case of $a_s <0$, the model Jost-Kohn potential $V_{-}(r)$ has only two parameters, $r_0$ and $a_s$ and so it is straightforward to model a Feshbach resonance by using these two parameters as fitting parameters provided the Feshbach-resonance phase shift $\eta_r$ admits an effective range expansion with $r_{f} > 0 $. Then $r_0 = r_{f}$ and the Feshbach-resonant scattering length is the same as in $V_{-}(r)$. In fact, even in the case of a narrow Feshbach resonance, $r_{f}$ is found to be positive in the regime of negative $a_{s}$ in many of ultracold atomic species \cite{pra:2013:hulet}. However, in the case $a_s >0$, we need three parameters, the third parameter $\kappa$ determines the binding energy of the bound state of an equivalent potential. In order to discuss how to model Feshbach-resonant interaction with $V_{+}(r)$, it may be instructive to recall the salient features of two-channel model of Feshbach resonances \cite{twochannel}  which has found considerable applications in modeling  magnetic Feshbach resonances (MFR) of ultracold atoms \cite{chin:rmp2010}.

In the two-channel model, the lower channel is open and the upper channel is closed, meaning that the asymptotic collision energy is above the threshold of the open channel but below the threshold of the closed channel. The closed channel is assumed to support a bound state $\psi_c(r)$ with binding energy $E_c$. There is a coupling $W(r)$ between the two channels. The $S$-matrix for the open channel is given by 
\bea 
S(k) =\exp[2 i \eta_{bg}(k)] \frac{E - E_c - E_{{\rm shift}} - i \hbar \Gamma_f/2} {E - E_c - E_{{\rm shift}} + i \hbar \Gamma_f/2}
\eea 
where $\eta_{bg}$ is a non-resonant background phase shift, $E_{{\rm shift}}$ is a shift of the closed-channel bound state due to its coupling with the bare scattering state of the open channel and $\Gamma_f$ is the Feshbach resonance width defined by 
$
\Gamma_f = 2 \pi \mid \int \psi_c(r) W(r) \psi_E(r) d r  \mid^2 
$  
where $\psi_E(r)$ is the bare scattering state of the open channel at collision energy $E$. The resonance phase shift $\eta_r$ is given by 
\bea 
\cot \eta_r = - \frac{E - \tilde{E}_c }{\hbar \Gamma_f/2}
\eea
where $\tilde{E}_c =  E_c + E_{{\rm shift}}$. The total phase shift is $\eta  = \eta_{bg} + \eta_r$. Therefore, we have 
\bea 
\cot \eta = \frac{\cot \eta_{bg} \cot\eta_r  - 1}{\cot \eta_{bg} + \cot \eta_r} 
\label{coteta}
\eea
Considering $\eta_{bg}$ being small, one may approximate $\cot \eta_{bg} \simeq - 1/k a_{bg}$ where $a_{bg}$ is the background scattering length. This approximation is particularly good for a narrow resonance. At low energy, $\Gamma_f$ may be proportional to $k$, because the energy-normalized scattering wave function asymptotically behaves as   $\psi_E(r) \sim \sqrt{k} r $. This will happen if $W(r)$ is most prominent beyond the range of the open channel potential.  Under these conditions, we may write 
$ 
k \cot \eta_r \simeq - \frac{1}{a_r} + \frac{1}{2} r_r k^2
$ 
where 
$
\frac{1}{a_r} = - \lim_{k\rightarrow 0} \frac{ 2 k \tilde{E}_c}{\hbar \Gamma_f}
$
 and 
$
r_r = -\lim_{k\rightarrow 0} \frac{2 \hbar k}{\mu \Gamma_f}
$.
 In an MFR, the energy $\tilde{E}_c$ of the quasi-bound state is magnetically tuned across the threshold of the open channel. The resonant scattering length $a_{r}$ is negative (positive) if  $\tilde{E}_c$ is positive (negative). The resonance at zero energy occurs when  $\tilde{E}_c$ is zero, in which case $a_r \rightarrow \infty$.

From Eq. (\ref{coteta}), we can obtain $k \cot \eta  \simeq - 1/\bar{a_s} + (1/2) r_{f} k^2 $ where 
$
\frac{1}{\bar{a_s}} = \frac{1}{a_r} \left ( 1 - \frac{a_{bg}}{a_r} \right )
$ and 
\bea 
r_{f} = 2 a_{bg}\left ( 1 + \frac{r_r}{a_r} - \frac{a_{bg}}{a_r} \right )  + r_r
\label{effectiver}
\eea
Close to resonance, $|a_{bg}| <\!< |a_r|$. Writing $\Gamma_f \simeq k G$, where $G$ is a constant, we note that the parameter $r_{r}$ is inversely proportional to $G$. Therefore, $r_r$ will be large for small $G$ or for a narrow resonance. Let us now assume that $a_{bg} >0$ and
consider the case $a_r > 0$. For a broad resonance, usually $r_{f}$ is positive and small. This means that, for a broad resonance $\kappa$ should be large and we may set $\bar{r}_0 \simeq r_0 = r_{f}$. In the case of far-off resonance, we may set $\kappa \rightarrow \infty$ and $r_0 = R_e$. In the case of a narrow resonance, the parameter $\kappa$ can be used to control the deviation of $r_{f}$ from $R_e$.  For a magnetic field very close to resonant magnetic field at which $a_s \rightarrow + \infty$, we can safely assume that $\kappa a_s >\!> 2$ so that $\bar{a}_s \simeq a_s $ as follows from Eq.(\ref{eq32}). Then the value of $\kappa$ can be set by equating Eq.(\ref{eq33}) with the experimentally observed $r_f$ assuming $r_0 = R_e$. The negative effective range can be mimicked by making  $\kappa r_0 < 4 $. In the universality regime \cite{chin:rmp2010}, $\bar a_{s}=1/\kappa$, implying that for $a_{bg}>0$ and $a_{bg}<<a_{r}$, $\Gamma_f \propto k\kappa$ which will happen if the bound state behaves as $e^{-kr}$. This means that inter-channel coupling should predominantly occur in the asymptotic limit of the bound state. Then $r_r\sim 1/\kappa$, therefore in the limit $\kappa\rightarrow 0$, $r_{r}$ will be large and the effective range $r_{f}$ as given by  Eq.(\ref{effectiver}) will be large negative indicating the breakdown of the effective range expansion. Universality regime is found to occur mostly in broad Feshbach resonances for which it has been shown that nonlinear energy dependence of the phase shift even very close to zero energy becomes important, suggesting that effective range expansion may fail at the universality regime \cite{efferangetheo}.

\section{Conclusions}\label{4} 
The foregoing analysis on the Jost-Kohn potentials reveal that these potentials can account for collision physics near zero-energy resonances with finite-range effects as can be exhibited by Feshbach resonances of ultracold atoms in certain physical situations. We have shown that the finite-range effects displayed by the phase shift corresponding to the potential $V_{+}(r)$ critically depends on the relative strength of the parameters $\kappa r_0$ and $\kappa a_s$, where $r_0>0$ and $a_s$ correspond to an equivalent potential with $\kappa \rightarrow \infty$. However, in the limit $\kappa \rightarrow 0$ the effective range and the scattering length obtained by effective range expansion of the phase shift are drastically modified. We have shown that the physical origin of the modification can be identified with a zero-energy resonance. A zero-energy resonance for $s$-wave occurs when the potential is about to support a bound state. By construction, $V_{+}(r)$ does not support any bound state but has  parametric dependence on the bound state energy. 
For $\kappa r_0 <\!<1$ and $\kappa a_s > 2$, the modified effective range $\bar{r}_0$ is found to  be negative although the modified scattering length $\bar{a}_s$ remains positive. On the other hand, $\bar{r}_0$ is positive for $\kappa a_s <\!< 1$. In both cases $\bar{r}_0$ can become quite large if  $\kappa r_0 <\!<1$. Thus, according to the theory of Jost and Kohn, it is possible to construct  a model potential which has no bound state but can provide the resonant scattering effects which may be induced by bringing a bound state close to the threshold of the potential as in the case of Feshbach resonances. We have also shown that $V_{+}(r)$ can describe universality regime where $a_s = 1/\kappa$ but then  $\bar{r}_0$ becomes negative and large. 

Under appropriate limiting conditions, as mention earlier, Jost-Kohn potentials reduce to P\"{o}schl-Teller form which has been extensively used for quantum Monte Carlo simulation of many body effects of a Fermi gas of atoms for $a_{s}\rightarrow-\infty$ \cite{carlson:prl:2003,forbes:pra86:2012,li:pra84:2011,morris:pra81:2010,schonenberg:pra:95:2017}. Therefore, the use of Jost-Kohn potentials in quantum simulation will open a broad perspective of $s$-wave many body physics of atomic gases. For a homogeneous many-particle system, many-body theories are conveniently developed in momentum-space under second quantization formalism where a model pseudo-potential with energy- or momentum-dependent phase shift may be applicable in order to explore effective range effects. However, for inhomogeneous systems like tightly confined atomic gases, such momentum-dependent description is not appropriate. So, Jost-Kohn potentials discussed in this paper will be particularly useful for developing many-body physics of trapped atomic gases. 

\vspace{0.5cm}
\begin{center}
{\bf Acknowledgments}
\end{center}
Two of us (SM and DS) are thankful to the Council of Scientific and Industrial Research (CSIR), Govt. of India, for support. KA and BD thankfully acknowledge the support from the Department of Science \& Technology, Govt. of India, under the project No. SB/S2/LOP-008/2014.

\appendix
\section{Derivation of negative-$a_s$ potential}
Now the first order potential $V_1(r)$ is evaluated as 
\begin{widetext}
\begin{eqnarray}
\lambda rV_1(r) &=& \frac{1}{2\pi i}\int_{-i\infty}^{i\infty}e^{rz}\frac{d}{dz}z[ g(z)-1]dz \nonumber \\
                &=& 2a^2\lambda r e^{-ar}
\end{eqnarray}
\end{widetext}
or
\begin{equation}
 V_1(r) = 2a^2 e^{-ar}
\end{equation}
\begin{widetext}
\begin{eqnarray}
 \int_0^\infty e^{-zr}V_m(r)dr &=& \sum_{l=2}^m \sum_{\nu_1+\nu_2+...+\nu_l=m} \int_0^\infty dr_1 \int_{r_1}^\infty dr_2...\int_{r_{l-1}}^\infty dr_l(1-e^{-zr_1})\frac{1}{z}(1-e^{-z(r_2-r_1)}) \nonumber \\&...\times&
 \frac{1}{z}(1-e^{-z(r_l-r_{l-1})})V_{\nu_1}(r_1)V_{\nu_2}(r_2)...V_{\nu_l}(r_l)
 \label{39}
\end{eqnarray}
\end{widetext}
Next two higher order potentials are evaluated as from the above equation
\begin{equation}
 \int_{0}^{\infty}e^{-zr}V_2(r)dr=2a^2\Big(-\frac{1}{z+a}+\frac{1}{z+2a}\Big)
\end{equation}
so that after the inverse Laplace transformation
\begin{equation}
 V_2(r)=2a^2(-e^{-ar}+2e^{-2ar})
\end{equation}
similarly
\begin{equation}
  V_3(r)=2a^2(e^{-ar}-4e^{-2ar}+3e^{-3ar})
\end{equation}
Finally, the whole potential is given by from Eq.(\ref{s1})
\begin{eqnarray}
V_{-}(r) &=& \lambda^1V_1(r)+\lambda^2V_2(r)+\lambda^3V_3(r)+...   \nonumber \\
     &=& 2a^2\Big[\lambda^1 e^{-ar}+\lambda^2 (-e^{-ar}+2e^{-2ar})+\lambda^3 (e^{-ar}+-4e^{-2ar}+3e^{-3ar})+...\Big]   \nonumber \\
     &=& 2a^2\Big[e^{-ar}(\lambda-\lambda^2+\lambda^3-...)+e^{-2ar}(2\lambda^2-4\lambda^3+...)+e^{-3ar}(3\lambda^3-...)+... \Big]  \nonumber \\
     &=& 2a^2\Big[\lambda e^{-ar} (1+\lambda)^{-1}+2\lambda^2e^{-2ar}(1+\lambda)^{-2}+3\lambda^3e^{-3ar}(1+\lambda)^{-3}+...\Big]   \nonumber \\
     &=& 2a^2e^{-ar}C\lambda[1+2\lambda Ce^{-ar}+3\lambda^2C^2e^{-2ar}+...] \hspace{.3in}{\rm where,}\hspace{.1in}C=(1+\lambda)     \nonumber \\
     &=& 2a^2e^{-ar}C\lambda[1-\lambda Ce^{-ar}]^{-2}  \nonumber \\
     &=& \frac{2a^2e^{-ar}\lambda(1+\lambda)^{-1}}{\big[1-\frac{\lambda}{1+\lambda}e^{-ar}\big]^2} \nonumber \\
     &=& \frac{2a^2e^{-ar}\lambda(1+\lambda)}{[1+\lambda(1-e^{-ar})]^2}
\end{eqnarray}

\section{Derivation of positive-$a_s$ potential}

Based on the Gel'fand-Levitan theory \cite{gelfand}, Jost and Kohn derived a 4-parameters potential $V(r)$ for $a_s>0$, where the four parameters are the $a_s$, $r_0$, $\kappa$ and the normalization constant $C$ of a single bound state assuming that the potential is capable of supporting only one bound state. 
The Jost function for $V(r)$ is 
\begin{equation}
 f(k)=\frac{4(k^2+\kappa^2)}{(2k+ib)(2k-ia)}
\end{equation}
So that the bound state is given by $k=-i\kappa$, which is a zero of $f(k)$.
\begin{widetext}
\begin{eqnarray}
 g(z)-1&=&\sum_{i=1}^m \int_0^\infty dr_1\int_{r_1}^\infty dr_2...\int_{r_l-1}^\infty dr_l \frac{1}{z^l}\times (1-e^{-zr_1})(1-e^{-z(r_2-r_1)})...(1-e^{-z(r_l-r_l-1)})V(r_1)\nonumber\\ \times V(r_2)&...&V(r_l)
\end{eqnarray}
\end{widetext}

The function $\bar g(z)$ is given from Eq.(\ref{eq26}) 
\begin{equation}
 \log \bar g(z) = \log \frac{(z+\xi_0)^2}{(z-b)(z+a)}
\end{equation}
We consider $\xi_{\pm}=\kappa\pm ik_r$ then
\begin{equation}
 \bar g(z)= \frac{(z+\xi_+)(z+\xi_-)}{(z+a)(z-b)} =\frac{(z+\kappa)^2+k_r^2}{(z+a)(z-b)} 
 \label{33}
\end{equation}
The explicit form of $V_+(r)$ is given by
\begin{equation}
 rV_1(r)=\frac{1}{2\pi i}\int_{-i\infty}^{i\infty}e^{rz}\frac{d}{dz}z[\bar g(z)-1]dz
\end{equation}
Now substitute $\bar g(z)$ in the above equation and we get
\begin{equation}
 V_1(r)= C_1 e^{-ar}+ C_2 e^{br}
\end{equation}
where $C_1=-\frac{(\kappa-a)^2+k_r^2}{1+\frac{b}{a}}$ and $C_2=-\frac{(\kappa+b)^2+k_r^2}{1+\frac{a}{b}}$.
With the reference of Eq.(\ref{39}), next higher order potential is given by
\begin{eqnarray}
V_2(r)=\Big(\frac{C_1e^{-ar}}{a}-\frac{C_2e^{br}}{b}\Big)^2-\frac{C_1^2e^{-ar}}{2a^2}-\frac{C_2^2e^{br}}{2b^2}+\frac{C_1C_2}{ab(a-b)}(ae^{-ar}-be^{br})
\end{eqnarray}
Similarly, higher order terms can be calculated.


\begin{thebibliography}{10}
\bibitem{JostKhon} R. Jost and W. Kohn, \href{https://doi.org/10.1103/PhysRev.87.977}{{Phys. Rev.} {\bf 84}, 977 (1952)}.
\bibitem{JostKhon1} R. Jost and W. Kohn, \href{http://gymarkiv.sdu.dk/MFM/kdvs/mfm\%2020-29/mfm-27-9.pdf}{{Dan. Mat. Fys. Medd} {\bf 27}, no. 9 (1953)}.
\bibitem{annphys:2008}E. Braaten, M. Kusunoki and D. Zhang, \href{https://doi.org/10.1016/j.aop.2007.12.004} {Ann. Phys. {\bf323}, 1770-1815 (2008)}.
\bibitem{bloch:2008} I. Bloch, J. Dalibard, and W. Zwerger, \href{https://doi.org/10.1103/RevModPhys.80.885}{{Rev. Mod. Phys.}{\bf 80}, 885 (2008)}.
\bibitem{chin:rmp2010} C. Chin, R. Grimm, E. Tiesinga, and P. S. Julienne,\href{https://doi.org/10.1103/RevModPhys.82.1225}{{Rev. Mod. Phys.} {\bf 82}, 1225 (2010)}.
\bibitem{sresonance} T. K\"{o}hler, K. Goral and P. S. Julienne, \href{https://doi.org/10.1103/RevModPhys.78.1311}{{Rev. Mod. Phys.} {\bf 78}, 1311-1361 (2006)}.
\bibitem{Huang1} T. D. Lee, K. Huang and C. N. Yang, \href{https://journals.aps.org/pr/pdf/10.1103/PhysRev.106.1135}{Phys. Rev. {\bf106}, 1135 (1957)}.
\bibitem{Huang2} K. Huang, \href{http://www.fulviofrisone.com/attachments/article/486/Huang,\%20Kerson\%20-\%201987\%20-\%20Statistical\%20Mechanics\%202Ed\%20(Wiley)(T)(506S).pdf}{{\it Statistical Mechanics}} (John Wiley \& Sons, 1987), Second edition.
\bibitem{efferangetheo} C. L. Blackley , P. S. Julienne and J. M. Hutson,\href{https://doi.org/10.1103/PhysRevA.89.042701}{{Phys. Rev. A} {\bf 89} 042701 (2014)}.
\bibitem{ohara:prl:2012} E. L. Hazlett, Y. Zhang, R. W. Stites and K.M.O'Hara,\href{https://doi.org/10.1103/PhysRevLett.108.045304}{{Phys. Rev. Lett}. {\bf 108}, 045304 (2012)}.
\bibitem{pra:2013:hulet} P. Dyke, S. E. Pollack, and R. G. Hulet, \href{https://doi.org/10.1103/PhysRevA.88.023625}{{Phys. Rev. A} {\bf 88}, 023625 (2013)}.
\bibitem{range} S. Roy, M. Landini, A. Trenkwalder, G. Semeghini, G. Spagnolli, A. Simoni, M. Fattori, M. Inguscio, and
G. Modugno, \href{https://doi.org/10.1103/PhysRevLett.111.053202}{{Phys. Rev. Lett}. {\bf 111}, 053202 (2013)}.
\bibitem{annphys:2012}C. Ji, D. R. Phillips and L. Platter, \href{https://doi.org/10.1016/j.aop.2012.02.001} {Ann. Phys. {\bf327}, 1803-1824 (2012)}.
\bibitem{efimov1} C. H. Greene, P. Giannakeas and J. P\'{e}rez-R\'{i}os, \href{https://doi.org/10.1103/RevModPhys.89.035006}{Rev. Mod. Phys.{\bf 89}, 035006 (2017)}.
\bibitem{efimov2} A. Kievsky, M. Gattobigio, \href{http://dx.doi.org/10.1103/PhysRevA.92.062715}{Phys. Rev. A {\bf92}, 062715 (2015)}.
\bibitem{efimov3} C. L. Blackley, C. R. L. Sueur, J. M. Hutson, D. J. McCarron, M. P. K\"{o}ppinger, H. W. Cho, D. L. Jenkin and S. L. Cornish \href{http://dx.doi.org/10.1103/PhysRevA.87.033611}{Phys. Rev. A {\bf87}, 033611 (2013)}.
\bibitem{Bolda_02} E. L. Bolda, E. Tiesinga and P. S. Julienne, \href{https://doi.org/10.1103/PhysRevA.66.013403} {Phys. Rev. A {\bf66}, 013403 (2002)}.
\bibitem{gelfand} I. M. Gel'fand and B. M. Levitan, { Dokl. Akad. Nauk. USSR} {\bf 77}, 557-560 (1951); I. M. Gel'fand and B. M. Levitan, {Izv. Akad. Nauk. SSRser. Math.} {\bf 15}, 309-360 (1951).
\bibitem{schieff_book} L. I. Schiff, {\it Quantum Mechanics} (Tata McGraw-Hill, 1968), Third edition.
\bibitem{LandauLifshitz_QM} L. D. Landau and E. M. Lifshitz, \href{http://power1.pc.uec.ac.jp/~toru/notes/LandauLifshitz-QuantumMechanics.pdf}{{\it Quantum Mechanics} (Pergamon Press, 1965), Vol. 3, Second edition}.
\bibitem{theory-frmodel} S. Z\"{o}llner, H. D. Meyer and P. Schmelcher, \href{https://doi.org/10.1103/PhysRevA.74.053612}{Phys. Rev. A {\bf74}, 053612 (2006)};\href{http://dx.doi.org/10.1103/PhysRevA.75.043608}{Phys. Rev. A {\bf75}, 043608 (2007)}; H. Wu and J. E. Thomas, \href{https://doi.org/10.1103/PhysRevA.86.063625}{{Phys. Rev. A} {\bf 86}, 063625 (2012)}.
\bibitem{fr1} P. I. Schneider, S. Grishkevich and A. Saenz, \href{https://doi.org/10.1103/PhysRevA.80.013404}{{Phys. Rev. A} {\bf 80}, 013404 (2009)}.
\bibitem{chin:pra:2009} A. D. Lange, K. Pilch, A. Prantner, F. Ferlaino, B. Engeser, H.-C. Nägerl, R. Grimm, and C. Chin, \href{https://doi.org/10.1103/PhysRevA.79.013622}{{Phys. Rev. A} {\bf 79}, 013622 (2009)}.
\bibitem{Flambaum} V. V. Flambaum, F. G. Gribakin and C. Harabati, \href{https://doi.org/10.1103/PhysRevA.59.1998}{{Phys. Rev. A} {\bf 59} 1998 (1999)}.
\bibitem{Ketterle:pra:2014} H. Veksler, S. Fishman and W. Ketterle \href{https://doi.org/10.1103/PhysRevA.90.023620}{{Phys. Rev. A} {\bf 90}, 023620 (2014)}.
\bibitem{twochannel} T. K\"{o}hler, T. Gasenzer, P. S. Julienne and K. Burnett,\href{https://doi.org/10.1103/PhysRevLett.91.230401}{{Phys. Rev. Lett.} {\bf 91}, 230401 (2003)}.
\bibitem{gao:jpb:2003} B. Gao \href{http://iopscience.iop.org/0953-4075/36/10/319} {J. Phys. B: At. Mol. Opt. Phys. {\bf36}, 2111 (2003)}.
\bibitem{gao:pra:2001} B. Gao \href{https://doi.org/10.1103/PhysRevA.64.010701} {Phys. Rev. A} {\bf64} 010701 (R) (2001).
\bibitem{bargmann}V. Bargmann, \href{https://doi.org/10.1103/RevModPhys.21.488}{{Rev. of Mod. Phys.} {\bf 21}, 488 (1949)}.
\bibitem{nume1} R. B. Johnson, \href{https://doi.org/10.1063/1.435384}{{J. Chem. Phys.} {\bf 67}, 4086 (1977)}.
\bibitem{bethe} H. A. Bethe, \href{https://doi.org/10.1103/PhysRev.76.38}{{Phys. Rev.} {\bf 76}, 38 (1949)}.
\bibitem{deb:injmp:2016} B. Deb,  \href{https://doi.org/10.1142/S0217979216500363}{{Int. J. Mod. Phys. B} {\bf 30}, 1650036 (2016)}.
\bibitem{carlson:prl:2003}J. Carlson, S. Y. Chang, V. R. Pandharipande and K. E. Schmidt, \href{https://doi.org/10.1103/PhysRevLett.91.050401} {Phys. Rev. Lett. {\bf91}, 050401 (2003)}.
\bibitem{forbes:pra86:2012}M. M. Forbes, S. Gandolfi and A. Gezerlis, \href{http://dx.doi.org/10.1103/PhysRevA.86.053603}{Phys. Rev. A {\bf86}, 053603 (2012)}.
\bibitem{li:pra84:2011}X. li, J. Kolorenč and L. Mitas, \href{https://doi.org/10.1103/PhysRevA.84.023615}{Phys. Rev. A {\bf84}, 023615 (2011)}.
\bibitem{morris:pra81:2010}A. J. Morris, P. L. R\'{i}os and R. J. Needs, \href{http://dx.doi.org/10.1103/PhysRevA.81.033619}{Phys. Rev. A {\bf81}, 033619 (2010)}.
\bibitem{schonenberg:pra:95:2017} L. M. Schonenberg and G. J. Conduit \href{https://doi.org/10.1103/PhysRevA.95.013633} {Phys. Rev. A {\bf 95}, 013633 (2017)}.
\bibitem{butsch:foundphys:1998}T. Busch, B. G. Englert, K. Rzazewski and M. Wilkens,\href{https://link.springer.com/article/10.1023/A:1018705520999}{{Foundation of Physics} {\bf 28}, 549 (1998)}.
\bibitem{studies} E. Tiesinga, C. J. Williams, F. H. Miles and P. S. Julienne, \href{https://journals.aps.org/pra/pdf/10.1103/PhysRevA.61.063416}{Phys. Rev. A {\bf61}, 063416 (2000)}.
\bibitem{physscrpt:partha} P. Goswami and B. Deb, \href{https://doi.org/10.1088/0031-8949/91/8/085401}{{Phys. Scr.} {\bf 91} , 085401 (2016)}.
\bibitem{fano1961} U. Fano, \href{https://doi.org/10.1103/PhysRev.124.1866}{Phys. Rev. {\bf124}, 1866 (1961)}. 
\bibitem{debandgsa:jpb:2009} B. Deb and G. S. Agarwal, \href{https://doi.org/10.1088/0953-4075/42/21/215203}{J. Phys. B: At. Mol. Opt. Phys. {\bf 42} , 215203 (2009)}.
\bibitem{vanderwaal} G. F. Gribakin and V. V. Flambaum, \href{https://doi.org/10.1103/PhysRevA.48.546}{{Phys. Rev. A} {\bf 48}, 546 (1993)}.
\bibitem{gao} B. Gao \href{https://doi.org/10.1103/PhysRevA.58.4222}{{Phys. Rev. A} {\bf 58} 4222 (1998)}.
\bibitem{effrange}T. L. Ho, X. Cui and W. Li, \href{https://doi.org/10.1103/PhysRevLett.108.250401}{{Phys. Rev. Lett}. {\bf 108}, 250401 (2012)}.




\end{thebibliography}
\end {document}